# Propagation of Uncertainty in Risk Analysis and Safety Integrity Level Composition

Jens Braband[1], Hendrik Schäbe[2]

[1]Siemens AG, Braunschweig, Germany
jens.braband@siemens.com
[2]TÜV Rheinland, Cologne, Germany
schaebe@de.tuv.com

**Abstract.** In many risk analyses the results are only given as mean values and often the input data are also mean values. However the required accuracy of the result is often an interval of values e. g. for the derivation of a Safety Integrity Level (SIL). In this paper we reason what should be the accuracy of the input data of risk analyses if a particular certainty of the result is demanded. Also the backside of the coin, the SIL composition is discussed. The results show that common methods for risk analysis are faulty and that SIL allocation by a kind of SIL calculus seems infeasible without additional requirements on the composed components. A justification of a common practice for parameter scaling in well-constructed semi-quantitative risk analysis is also provided.

**Keywords:** risk analysis, uncertainty, error propagation, safety integrity level.

## 1 Introduction

In quantitative risk analysis, the formulae used are usually composed of two operations, namely summation and multiplication. The complete risk for a project is usually composed as the risk of individual contributions, e.g. from different hazards, while each contribution can often be written as products, e.g. the product of a frequency, risk reduction factors and severity. So, in many cases, the overall risk can be expressed as a sum of products as in the following example [1]:

$$IRF_i = N_i \sum_{\text{all hazards } H_j} HR_j (D_j + E_{ij}) \sum_{\text{all accidents } A_k} C_{jk} \times F_{ik}$$

There is no need to explain the details here; it is only important to understand that, while all parameters are random variables, commonly such formulae are only evaluated by their means, not taking into account any uncertainty. In some published cases, the sum extends over up to 50 hazards with typically about five accident types. In fact, the second sum represents the results of an event tree analysis, where the C parameter is typically a product of three risk reduction factors.

Risk must not be confused with uncertainty, as is often done in financial contexts. An excellent discussion is given by [2], where it is proposed that risk should be judged not only by the hazards that can happen and their associated consequences, but also by the uncertainty about the occurrence of the hazards and the severity of the consequences. In this paper, we consider the statistical uncertainty of the parameters of a risk analysis only, knowing that other uncertainties also exist, e.g. about the knowledge or the model itself.

Uncertainty alone cannot be used as a definition of risk, as large uncertainties need only to be regarded when the frequency of the hazards or the severity of the consequence is high. It is only if both parameters are more or less fixed that the remaining risk depends only on the uncertainty.

A major goal in risk analysis is to derive a tolerable hazard rate (THR) or probability of dangerous failure per hour (PdfH) $H$ that falls into a particular safety integrity level (SIL) with a high degree of certainty

$$P(H \in [10^{-(x+1)}, 10^{-x}]) = 1 - \alpha \qquad (1)$$

In order to achieve this, we could choose an estimate $\mu_H$ for the THR so that

$$\mu_H \pm q_\alpha \sigma_H \qquad (2)$$

falls completely into the SIL range. This would be an approach similar to the famous $2\sigma$ or $3\sigma$ rules of thumb for the normal distribution, which correspond to more than 95% and 99% confidence, respectively. Having fixed $q_\alpha$, our measure of uncertainty in this paper is given by $\sigma_H$.

A similar problem occurs, when systems shall be composed of subsystems of a lower SIL [7]. While some standards such as IEC 61508 [5] or ISO 26262 [6] contain such composition rules, the big problem is that they are conflicting. Not only do different standards propose different rules, but sometimes there are also contradictions within single standards. E. g. IEC 61508 allows SIL composition for SW components (called systematic capability in clause 7.4.3.2 of part 2), but only once. So a SIL 3 component may be implemented by two SIL 2 subcomponents but this process must not be iterated. However in section 7.4.4 of the same standard a general decomposition is allowed with respect to HW fault tolerance. And finally ISO 26262 in part 9 allows a similar general decomposition.

A closer look at SIL composition reveals that the process and the calculations involved are quite similar to those in risk analyses (compare for example figure 6 of IEC 61508-2). Usually the combination rules are logical AND and OR, which under common probabilistic approaches such as those used in fault tree analysis translates to summation and multiplication of the safety measures translates to a sum of products as in formula (1).

## 2  Normal Distributions

For independent normal random variables, the variance of the sum equals the sum of the variances. Given unbiased estimates and variances of *n* normal contributions of a similar magnitude, in order to achieve a particular certainty of the sum, the certainty of the contributions must be more accurate by a factor of $\sqrt{n}$. So, given ten contributors, each contribution would have to be about three times more certain than the final result. This can be quite demanding and could be hard to achieve for, say, 100 contributions. But even for the sum X+Y, the increase in the standard deviation would be 41%. It should be noted that the increase slows down to 73% and 100% for two and three operations.

Distribution of the product of normal random variables is less simple. We start with a product of two independent normal variables, say X and Y. It is easy to derive that, for Z=XY,

$$\sigma_Z^2 = \sigma_X^2 \sigma_Y^2 + \mu_X^2 \sigma_Y^2 + \mu_Y^2 \sigma_X^2 \qquad (3)$$

holds. We illustrate the results by a first example. Assume we aim at a high certainty of the final result by applying the $3\sigma$ rule. We take a hazard rate H satisfying (1) and (2) by

$$\mu_X = 0.55 \cdot 10^{-x} \text{ and } \sigma_X = 0.15 \cdot 10^{-x}$$

This corresponds to a safety requirement of SIL x. We now assume a risk reduction factor with the parameters

$$\mu_Y = 10^{-y} \text{ and } \sigma_X = 0.3 \cdot 10^{-y}.$$

We now apply (3) which leads to $\sigma_Z \sim 0.23 \cdot 10^{-(x+y)}$. We have to bear in mind that $\mu_Z = 0.55 \cdot 10^{-(x+y)}$, so that there is a relative increase in the standard deviation by about 50%. This is in the same order of magnitude as for the sum. Figure 1 shows a plot of a kernel density estimator (10,000 samples) for Z and a normal curve with the same parameters. For comparison, a normal distribution with the same standard deviation as for H is also plotted (points only) in order to illustrate the increase in variance.

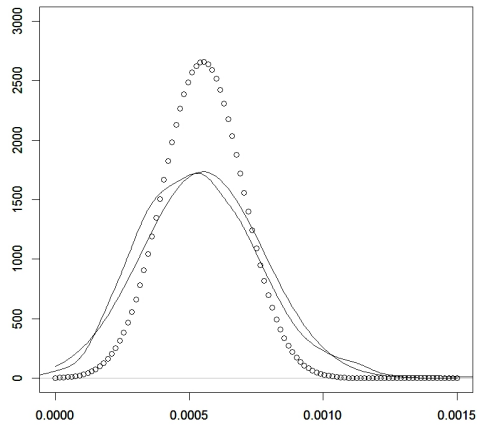

**Fig. 1**: Increase in variance for the product XY

We consider a second case, the "halving" of SIL intervals as is often applied in semi-quantitative risk analysis, e.g. [4]. Here, the mean values for the parameter estimates are $\sqrt{10} \cdot 10^{-x}$ and $10^{-x}$, respectively. Each hazard is treated separately and is compared to a risk acceptance criterion, so that each risk is estimated by a hazard rate (HR) $H$, which may be reduced by one or more risk reduction factors or barriers $B_i$ multiplied by a severity $S$. In practice, if only one barrier exists, $H \cdot B \cdot S$ represents a common case. We take

$$\mu_H = 0.55 \cdot 10^{-x} \text{ and } \sigma_H = 0.015 \cdot 10^{-x}$$
$$\mu_B = \sqrt{10} \cdot 10^{-y} \text{ and } \sigma_B = 10^{-y}$$
$$\mu_S = 10^s \text{ and } \sigma_X = \sqrt{10} \cdot 10^{s-1}$$

as some obvious choices. Figure 3 again shows the sample distribution, its normal approximation and the optimal normal curve, which has the same ratio of mean to standard deviation as H. It is important to note the skewness of the sample distribution and that the relative increase in standard deviation is about 100%.

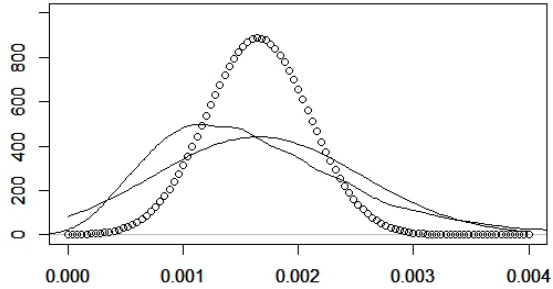

**Fig. 2**: Increase in variance for the risk (x=1, y=2, s=0)

This example clearly illustrates that, in order to achieve the target certainty of the result, the certainty of the input parameters must be much better, depending on the number of parameters. This justifies that, for semi-quantitative analysis, the bandwidth of the input classes should be smaller than the bandwidth of the results. If the bandwidth of the results is of an order of magnitude as in SIL scaling, then the choice of a bandwidth with a scaling factor of $\sqrt{10}$ seems appropriate.

On the other hand, we could fix the target uncertainty, e.g. $\sigma_Z = 0.15 \cdot 10^{-z}$. Assuming $\mu_Z = \mu_X \mu_Y$ and $\sigma_X = \sigma_Y$, we may derive a quadratic equation for $\sigma_X^2$

$$(\sigma_X^2)^2 + (\mu_X^2 + \mu_Y^2)\sigma_X^2 - \sigma_Z^2 \leq 0.$$

## 3 Log-normal Distributions

In order to treat a product of two random variables in an analytically simple way, it would be desirable if the random variables X and Y as well as their product Z followed the same family of distributions. Taking the logarithm of both sides of the definition of Z, we arrive at

$$\ln(Z) = \ln(X) + \ln(Y). \qquad (4)$$

We are now looking for a family of distributions that is valid for random variables and their sums. This is just the problem of so-called infinitely divisible distributions [3]. This means that, if we choose an infinitely divisible family of distributions for ln(X), all computations can be carried out with the same parametric model. An infinitely divisible distribution f(x) is characterized by its characteristic function [3]

$$\phi(t) \stackrel{\text{def}}{=} \int_{-\infty}^{\infty} f(x) \exp\{ixt\}\, dx = \exp\left\{i\alpha t + \int_{-\infty}^{\infty} \frac{\exp(itx) - 1 - itx}{x^2} dK(x)\right\} \quad (5)$$

This formula describes the characteristic function of an infinitely divisible random variable with finite dispersion. Here, K(.) denotes a non-decreasing and bounded function. If a parametric model that is described by (5) is now chosen, there are two well-known families of distributions:

a) gamma distributions,
b) normal distributions.

The gamma distribution family, however, requires that all distributions have the same scale parameter. This would mean a considerable restriction for the problem that we want to treat. There is a second argument for the normal distribution. Considering the logarithmic version of (1)

$$P(\log(H) \in [-x-1; -x]) = 1 - \alpha \qquad (6)$$

where log denotes the decimal logarithm, we see that the problem as such admits logarithmic scaling. It is then natural to assume normally distributed errors on top of this scaling.

If the characteristics X' = ln(X), Y' = ln(Y) and Z' = ln(Z) now have normal distributions with parameters $(\mu'_X, \sigma'_X), (\mu'_Y, \sigma'_Y)$ and $(\mu'_Z, \sigma'_Z)$, respectively, they are connected by the following equations:

$$\mu'_Z = \mu'_X + \mu'_Y,$$
$$\sigma'^2_Z = \sigma'^2_X + \sigma'^2_Y.$$

The mean and variance of the random variables X, Y and Z are computed from the parameters by the following equation:

$$\mu = (\mu' + \sigma'2/2)$$
$$\sigma^2 = \exp(2\mu' + \sigma'^2)(\exp(\sigma'^2) - 1)$$

where the indices X, Y or Z have been omitted. For the example from the previous section, the following computation is carried out. From the mean and standard deviation of X

$$\mu_X = 0.55 \cdot 10^{-x} \text{ and } \sigma_X = 0.15 \cdot 10^{-x}$$

the parameters

$$\mu'_X = -0.6337 + ln(10^{-x}), \sigma'_X = 0.2679$$

of the log-normal distributions are obtained. For Y, with a mean and standard deviation

$$\mu_Y = 10^{-y} \text{ and } \sigma_X = 0.3 \cdot 10^{-y}.$$

the results are

$$\mu'_Y = -0.0431 + ln(10^{-y}), \sigma'_X = 0.2936$$

Combining Z=XY, the parameters for Z are

$$\mu'_Z = -0.6768 + ln(10^{-(x+y)}), \sigma'_Z = 0.3974$$

which gives the mean and standard deviation of the combined factor

$$\mu_Z = 0.55 \ 10^{-(x+y)},$$

$$\sigma_Z = 0.2275 \ 10^{-(x+y)}.$$

Remember that, in the previous section, with a normal distribution almost the same results have been computed. The value for $\mu_Z$ is identical; the value for $\sigma_Z$ is quite close to the value obtained above. The following figure shows the density of the log-normal distribution together with the curves already shown in Figure 1.

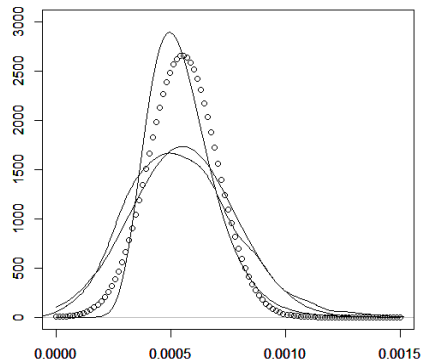

**Fig. 3:**   Density of the log-normal distribution of Z together with the simulation of XY and the distributions of X and Y

This density is slightly skewed. In any case, from the data used, it can be seen that the variance of Z is larger than that of X and Y. This repeats just the result from the previous section. It can also be seen that the log-normal density is close to the simulated density of XY, although not identical. Therefore, the log-normal distribution can be used as an approximation to analytically derive the similar results as with the methods applied to normally distributed errors.

## 4  Conclusions

In this paper, we have shown that, for the two common operations in quantitative risk analysis and SIL combination, summation and multiplication, uncertainties in the parameters of the input parameters lead to increasing uncertainties of the result.

We have treated two cases explicitly: normally and log-normally distributed parameters. In both cases, we have observed a substantially larger statistical uncertainty of the result, compared with the input factors. It also seems that the increase is more or less the same for both distributions. As a rule of thumb, we propose to reckon with an increase of about 50% for one operation, and an additional increase of about 30% for each additional operation.

This also leads to a justification of a common practice in well-constructed semi-quantitative risk analysis, which is to choose a smaller bandwidth in the parameters than the required bandwidth of the result. So, for the most common choice of a scaling factor of $\sqrt{10}$ for the input and 10 for the results, it would be admitted at least to double the standard deviation, which would allow the use two or three operations on average, this fitting in well with common practice.

These results also apply to the calibration of well known semi-quatitative risk analysis methods proposed by standards, e. g. the risk graph in IEC 61508 or the risk matrix in ISO 26262. Our results indicate that both methods in the standards are not well constructed. It seems as if the input parameter are on a decadic interval scale, meaning that the results don't fit into the SIL schematic, if uncertainties are properly accounted for.

The same observation holds for SIL composition. If several components would filful a SIl x, then any combination, be it summation or multiplication, would generally not be compatible with a target SIL x+1. So, unless we take uncertainty properly into account, all SIL combination rules proposed so far are not consistent with the SIL definition. In practice this would mean that components which shall be used to fulfill a mored demanding SIL in combination, must fulfill more stringent requirements than the ordinary SIL.